\documentclass{article}
\usepackage{spconf,amsmath,graphicx,amssymb}
\usepackage{float}
\usepackage{booktabs}
\usepackage{enumitem}
\usepackage{bbding}
\usepackage{pifont}
\usepackage{wasysym}
\usepackage{utfsym}

\title{Dance2MIDI: Dance-driven Multi-Instruments Music Generation}
\name{Bo Han$^{1}$, Yuheng Li$^{1}$, Yixuan Shen$^{2}$, Yi Ren$^{3}$, and Feilin Han$^{4}$ \thanks{Project page: https://dance2midi.github.io/}}
\address{\textsuperscript{1}College of Computer Science and Technology, Zhejiang University\\
\textsuperscript{2}National University Singapore\\
\textsuperscript{3}Bytedance AI Lab\\
\textsuperscript{4}Department of Film and TV Technology, Beijing Film Academy}
\begin{document}
%
\maketitle
\begin{abstract}
Dance-driven music generation aims to generate musical pieces conditioned on dance videos. Previous works focus on monophonic or raw audio generation, while the multi-instruments scenario is under-explored. The challenges associated with the dance-driven multi-instrument music (MIDI) generation are twofold: 1) no publicly available multi-instruments MIDI and video paired dataset and 2) the weak correlation between music and video. To tackle these challenges, we build the first multi-instruments MIDI and dance paired dataset (D2MIDI). Based on our proposed dataset, we introduce a multi-instruments MIDI generation framework (Dance2MIDI) conditioned on dance video. Specifically, 1) to capture the relationship between dance and music, we employ the Graph Convolutional Network to encode the dance motion. This allows us to extract features related to dance movement and dance style, 2) to generate a harmonious rhythm, we utilize a Transformer model to decode the drum track sequence, leveraging a cross-attention mechanism, and 3) we model the task of generating the remaining tracks based on the drum track as a sequence understanding and completion task. A BERT-like model is employed to comprehend the context of the entire music piece through self-supervised learning. We evaluate the generated music of our framework trained on the D2MIDI dataset and demonstrate that our method achieves State-of-the-Art performance.
\end{abstract}

\section{Introduction}
\label{sec:intro}
As choreographer Zakharov puts it "\textit{Music is the soul of dance; music contains and determines the structure, characteristics, and temperament of dance}". The relationship between music and dance is complementary. Studies have demonstrated that humans utilize the same neural pathways to appreciate both dance and music~\cite{zhuge2004knowledge,mastroianni2005super}. Therefore, it is essential for the accompanying music that conform to the fundamental structure, style, and emotional expression to enhance the artistic appeal of the dance videos.

In the era of short videos, sharing dance performances on social media platforms has become a popular trend. Mainstream platforms often provide automatic soundtracks for dances or allow creators to independently select music clips. However, it should be noted that the music available in these libraries is pre-existing and may only be suitable for simple and regular dance movements. Matching complex and diverse movements can be challenging. Additionally, the use of pre-existing music may result in copyright disputes. Manually selecting appropriate music for dance can be a time-consuming process. To improve matching, originality, and efficiency, automatic music generation has emerged as a thriving subject of research in recent years~\cite{aggarwal2021dance2music,di2021video,gan2020foley,kao2020temporally,li2021ai,zhu2022quantized}.

While there has been significant research on music-to-dance generation~\cite{han2023amd,kim2022brand,lee2019dancing,li2022danceformer,siyao2022bailando}, the inverse direction of dance-to-music generation remains underexplored, which is a challenging task for the following reasons:
\begin{itemize}
    \item Music generation is challenging~\cite{briot2020deep,ji2020comprehensive}. In real-world applications, music is often polyphonic and multi-instrumental, requiring harmony and coherence across all instruments. This complexity in music representation makes the generation difficult.
    \item Conditional music generation is also a challenging task~\cite{aggarwal2021dance2music,di2021video,su2021does}. The correlation between the music and the control signals, such as dance videos, is often weak. For instance, this correlation may include musical and dancing beats, tempo, and emotion~\cite{wang2023songdriver2}. However, there are many degrees of freedom for each modality (music and dance), which can be regarded as noise and may confuse the generative model during training.
    \item The lack of publicly available datasets containing paired music and dance videos hinders the development of dance-to-music generation research.
\end{itemize}
There are only a few works studying dance-to-music generation: Dance2Music~\cite{aggarwal2021dance2music} takes in the local history of the dance similarity matrix as input and generates monophonic notes. However, the handcrafted features they used may discard much useful information in dance videos and monophonic music is not applicable to the real scenarios. D2M-GAN~\cite{zhu2022quantized} takes dance video frames and human body motions as input to directly generate music waveforms. While this approach can generate continuous multi-instrumental music, the high variability of waveform data (e.g., variable and high-dynamic phase, energy, and timbre of instruments) makes it difficult to directly model high-quality waveforms. As a result, the generated music often contains strange noise.

This work aims to tackle the challenges of dance-to-music generation and address the issues of previous works. To overcome the scarcity of datasets, we collect and annotate the first large-scale paired dataset of dance and multi-instrument music (D2MIDI), which encompasses six mainstream dance genres: classical, hip-hop, ballet, modern, latin, and house. In total, it contains 71,754 pairs of multi-instrument MIDI data and dance video data. 
To model the correlation between music and dance, we introduce a multi-instrument MIDI generation framework (Dance2MIDI). It is architecturally designed with three primary modules: a Context Encoder for understanding the dance motion features related to music, a Drum Rhythm Generator for creating a base rhythmic drum track, and a Multi-Track MIDI BERTGen for producing multi-track MIDI track based on the drum track.

In light of the diversity in human skeleton space and the intricacy of associated movement patterns, we represent the human skeleton as a motion graph. To augment the feature extraction efficacy of the Context Encoder, we employ a graph convolutional network~\cite{yan2018spatial}. Our approach involves bifurcating the process into two distinct branches, each dedicated to extracting specific features: one for dance movement and the other for dance style.
The drum, being fundamental to the establishment of rhythm in music, often serves as the starting point for composers when crafting a new musical piece. This is typically achieved by designing the rhythm for the accompanying drum track. In this context, we utilize a Transformer as the core of the Drum Rhythm Generator. This generator progressively creates drum tracks through autoregression, guided by dance condition control information.
For the generation of other audio tracks, considering the unique characteristics of symbolic music in sequence modeling, we reframe this task as a sequence comprehension and completion task. Consequently, we introduce a model akin to BERT~\cite{devlin2018bert} to understand the entire MIDI music sequence in a self-supervised manner.
Through experimentation, we have found that our method can achieve harmonious and coherent multi-instrument dance-to-music generation and outperforms all baseline methods~\cite{aggarwal2021dance2music,di2021video,zhu2022quantized}, demonstrating the effectiveness of our dataset and framework. 
In summary, our main contributions are as follows:

\begin{itemize}
    \item We construct the first multi-instrument dance-to-music dataset (D2MIDI), which facilitates research in the field of dance-to-symbolic music generation.
    \item We introduce an effective multi-instrument dance-to-multi track music framework Dance2MIDI, which demonstrates the feasibility of multi-instrument music generation and provides insights into multi-modal symbolic music generation.
    \item Exhaustive qualitative and quantitative assessment demonstrate that our method achieves State-of-the-Art performance.
\end{itemize}

\section{Background}
\subsection{Music Generation}
While the waveform is the original form of audio, some models generate audio directly in the waveform~\cite{donahue2018adversarial,goel2022s,oord2016wavenet}. However, a single second of audio waveform spans tens of thousands of timesteps. As a result, existing non-symbolic music-based generative methods typically employ intermediate audio representations for learning generative models~\cite{dhariwal2020jukebox,kumar2019melgan,vasquez2019melnet}. Nevertheless, this does not completely alleviate the dilemma~\cite{ji2020comprehensive}. 
Consequently, some recent works have adopted a symbolic music modeling approach. 
MuseGAN~\cite{dong2018musegan} employs a multi-track GAN-based model using 1D piano-roll symbolic representations. Music Transformer~\cite{huang2018music} generates long sequences of music using 2D event-based MIDI-like audio representations.
Despite the potential of generative models for long sequence generation, the quality of samples produced by these models often deteriorates significantly. To address this issue, TBPTT~\cite{muhamed2021symbolic} employs a Transformer-XL~\cite{huang2018music} generator in conjunction with a pre-trained Span-BERT~\cite{joshi2020spanbert} discriminator for long symbolic music generation, which enhances training stability.
PopMAG~\cite{ren2020popmag} proposes a novel Multi-track MIDI representation MuMIDI that enables simultaneous multi-track generation in a single sequence and introduces extra long-context as memory to capture long-term dependency in music.
SymphonyNet~\cite{liu2022symphony} introduces a novel Multi-track Multi-instrument representation that incorporates a 3-D positional embedding and a modified Byte Pair Encoding algorithm for music tokens. Additionally, the linear transformer decoder is employed as the backbone for modeling extra-long sequences of symphony tokens.

\subsection{Dance To Music}
A recent novel approach to dan\-ce beat tracking has been proposed~\cite{pedersoli2020dance}, which only detects music beats from dance videos. RhythmicNet~\cite{su2021does} employs a three-stage model comprising video2rhythm, rhythm2drum, and drum2music. However, it is limited to generating music for only two instruments. CMT~\cite{di2021video} establishes three relationships between video and music, including video timing and music beat, motion speed and simu-note density, and motion saliency and simu-note strength. While this approach does not specifically target dance-to-music tasks and fails to fully exploit the human motions present in dance videos. Dance2Music utilizes the local history of both dance similarity matrices to predict notes but is restricted to generating single-instrument music. D2M-GAN takes dance videos and human body motions as input to directly generate music waveforms. However, the generated music often contains noise.

\subsection{Symbolic Music Dataset}
The Groove MIDI Dataset (GMD)~\cite{gillick2019learning} comprises 13.6 hours of aligned MIDI and synthesized audio of human-performed, tempo-aligned expressive drumming, including 1,150 MIDI files and over 22,000 measures of drumming. 
In contrast, the Lakh MIDI dataset~\cite{raffel2016learning} is a collection of 176,581 unique MIDI files, with 45,129 matched and aligned to entries in the Million Song Dataset.
The MAESTRO dataset~\cite{hawthorne2018enabling} is a dataset composed of 198.7 piano MIDI, audio, and MIDI files aligned with 3 ms accuracy.
ADL Piano MIDI~\cite{ferreira_aiide_2020} is a dataset that is based on the Lakh MIDI dataset. It generates 9,021 pieces of piano MIDI data from the Lakh MIDI dataset and then crawls an additional 2,065 pieces of piano MIDI data from network channels. Both datasets contain only MIDI music of the piano instrument type. All the above datasets are purely symbolic and lack corresponding dance movement annotations. They can't support the task of dance-to-music.
The AIST Dance Video Database~\cite{tsuchida2019aist} provides a large-scale collection of dance videos with paired dance action videos and music annotations. However, its paired music is in the form of waveforms and lacks paired symbolic music annotations. Additionally, one piece of music corresponds to multiple videos, and the non-overlapping music clips comprise only 60 pieces.

\section{D2MIDI Dataset}
\begin{table*}[htb]
    \centering
    \begin{tabular}{c|c|c|c|c|c|c|c|c}
    \toprule
     Dataset  &Dance &Audio &MIDI   &Genres &Instrument &Pieces &Hours &Available\\ \midrule
     Groove MIDI  &\usym{2717} &\usym{2717} &\usym{2713} &- &1 &1,150  &13.6 &\usym{2713}\\
     LMD-aligned MIDI  &\usym{2717} &\usym{2713} &\usym{2713} &- &10 &45,129  &- &\usym{2713}\\
     MAESTRO Dataset  &\usym{2717} &\usym{2713} &\usym{2713} &1 &1 &1,276  &198.7 &\usym{2713} \\
     ADL Piano MIDI &\usym{2717} &\usym{2717} &\usym{2713} &- &10 &11,086  &- &\usym{2713} \\ \midrule
     \textit{AIST Database}  &\usym{2713} &\usym{2713} &\usym{2717} &10 &- &60  &118.1 &\usym{2713}\\ 
     Ours  &\usym{2713} &\usym{2713} &\usym{2713} &6 &13 &71,754 &\textbf{597.95} &\usym{2713}\\
     \bottomrule
    \end{tabular}
    \caption{Symbolic music dataset comparisons. The AIST Database does not contain symbolic music. However, it is often used in dance-to-music tasks for modeling non-symbolic music. In this paper, we also labeled it for comparison experiments.}
    \label{tab:dataset}
\end{table*}
\label{sec:D2MIDI}
In this section, we provide a brief overview of our newly collected dance-to-MIDI dataset (D2MIDI) and the methodology of its acquisition. D2MIDI represents the first multi-instrument dataset of its kind and possesses several notable features: 
\begin{itemize}
    \item High-quality solo dance video: it comprises high-quality solo dance videos that have been carefully curated from internet sources to 
    exclude low-quality footage and videos featuring multiple dancers (Section~\ref{subsec:craw})
    \item Multi-instrument and polyphonic MIDI: the dataset contains multi-instrumental and polyphonic MIDI transcriptions that are temporally synchronized with the corresponding dance videos (Section~\ref{subsec:transcrip})
    \item Multi-style and large-scale: the dataset is both multi-style and large-scale, encompassing a diverse range of dance styles across 71,754 clips. (Section~\ref{subsec:staistic}).
\end{itemize}

\subsection{Video Crawling and Selection}
\label{subsec:craw}
We manually filter dance videos from various video platforms using the following screening criteria: 1) The video must have a pure background with minimal interference from other characters. 2) Only videos featuring a single dancer are selected. 3) The music and dance movements must be highly synchronized. 4) The background music must be clear, and free of extraneous noise.

\subsection{MIDI Transcription and Annotation}
\label{subsec:transcrip}
To ensure consistency in the Frames Per Second (FPS) with the dance motions in the videos~\cite{siyao2022bailando}, we first standardize the FPS of all dance videos to 20. Additionally, we unify the sample rate of the audio to 10,240Hz and then separate the audio in the video.
Next, we utilize the MT3~\cite{gardner2021mt3} music transcription model to convert the original audio into multi-instrument MIDI music. However, the MIDI transcribed by MT3 may contain low-quality notes and discrepancies between music tempo changes and character movement changes in dance videos. To address these issues, we enlist professionals to align and label the MIDI music with reference to the context of the video and music. Specifically, professionals adjust the pitch, start time, duration, and instrument type of notes at corresponding positions in the music based on the pleasantness of the music and the context of the video.

\subsection{Dance Motion Estimation}
\label{subsec:motion_estimation}
The movement and posture of the human body are closely related to the fluctuations in the music. Unlike other 2d-keypoints methods~\cite{cao2017realtime,sun2019deep}, we extract the 3d keypoints of the human body including body, hand, and face, allowing densely represented pose features.

\subsection{Statistics}
\label{subsec:staistic}
We employ the sliding window method to sample data from the video. Each sampling window has a size of 600 frames, equivalent to a 30-second dance video, with a sliding window size of 40 frames. This process resulted in a total of 71,754 pairs of data, in which the dance type includes classical, hip-hop, ballet, modern, latin, and house. The music in each data pair does not repeat each other. In the D2MIDI dataset, the duration in each data pair is 30 seconds, which is guaranteed to generate music with a rhythmic structure.  The music in the pair contains up to 13 instrument types, including Acoustic Grand Piano, Celesta, Drawbar Organ, Acoustic Guitar (nylon), Acoustic Bass, Violin, String Ensemble 1, SynthBrass 1, Soprano Sax, Piccolo, Lead 1 (square), Pad 1 (new age) and Drum. We compare our proposed dataset with public datasets in Table~\ref{tab:dataset}.

\section{Dance2MIDI Framework}
\label{sec:format}

\begin{figure*}[htb]
    \centering
    \includegraphics[width=1\linewidth]{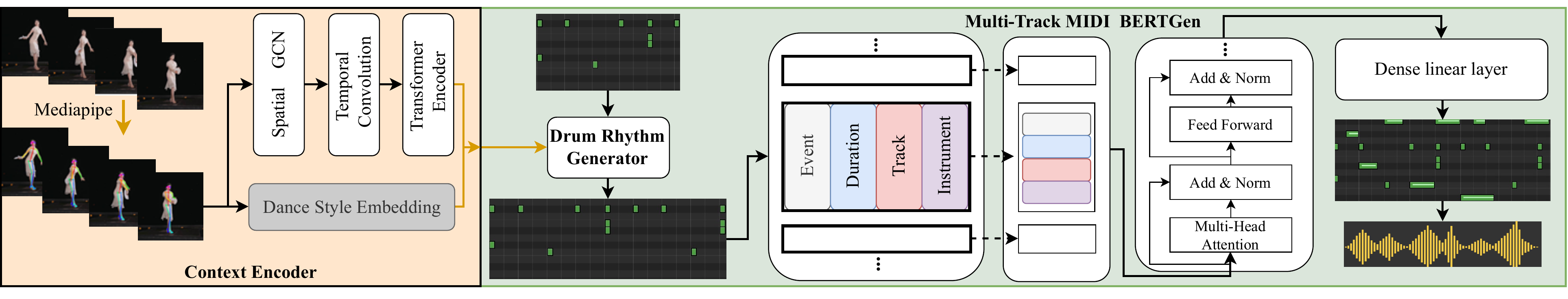}
    \caption{An overview of our proposed Dance2MIDI model. The dance video is input into the Mediapipe framework to extract the coordinates of the human body’s joint points. These coordinates are then used to encode the spatio-temporal features of dance movement and dance style (yellow block). Subsequently, these features serve as conditional information to guide the generation of multi-instrument MIDI music sequences (green block).}
    \label{fig:model_overview}
\end{figure*}

The proposed architecture is schematically illustrated in Fig.~\ref{fig:model_overview} and comprises three main components: the Context Encoder, the Drum Rhythm Generator, and the Multi-Track BERTGen. 
In the Context Encoder, we commence by employing the joint point extraction to obtain the spatial coordinates of the human joints within the dance video. Subsequently, via the utilization of two distinct branches, we extract the dance style features and dance movement features. These extracted features are then combined to form a concatenated representation, which serves as a guide for generating conditional control information that corresponds to the MIDI music. It is worth noting that in the realm of MIDI music, drums play a pivotal role in generating fundamental rhythm patterns that underlie the musical composition. Moreover, in the context of composition and improvisation, it is customary for composers to initiate the creation of a new musical piece by designing the rhythm for the accompanying drum track. As the piece progresses, additional instrumental tracks are incrementally layered on top of the drum track, thereby culminating in the production of the final musical composition. 
So we first leverage Drum Rhythm Generator to incrementally generate drum tracks in an autoregressive manner, thus establishing the foundational melody of the music. Subsequently, we augment the overall music composition by incorporating note information from other tracks and instruments, thereby enhancing its richness and complexity. We conceptualize this process as a sequence completion task, wherein the BERT-like model is employed to enrich the remaining music track, facilitating a comprehensive understanding of the entirety of the musical piece.

\subsection{Context Encoder}
The Context Encoder primarily comprises two branches designed to extract features related to dance movement and dance style. Initially, the human body motion joint coordinate $X \in \mathbb{R}^{T \times J \times 3}$ is extracted from the original dance video via the Mediapipe framework~\cite{lugaresi2019mediapipe}, which is then input into two conditional encoders.

In the dance movement branch, the human body motion joint is modeled as a motion graph. The spatial position of the character’s joints in each frame is first aggregated through a spatial Graph Convolutional Network (GCN). Subsequently, timing information across time frames is aggregated via temporal convolution. After this, we obtain the dance movement features $Z_m \in R^{T \times F}$, where $T$ and $F$ represent the number of video frames and the number of feature channels, respectively. 
It's particularly noteworthy that the relationship between dance movements and the beats of music is intricately linked, as it is the transitions in dance movements that often drive changes in musical beats. In this context, we transform the dance movement feature into a binary detection problem for music beats: Given the dance movement feature $Z_m \in R^{T \times F}$, motion information across both temporal and spatial dimensions is consolidated via attention learning within the Transformer encoder, yielding the final beat binary sequence $Z_b \in R^{T}$ of the same length as $Z_m$, where each frame is classified as either a beat or non-beat. This approach lays the foundation for the entire musical piece, ensuring consistency in timing and rhythm between the dance movement and the music.

The dance style branch operates similarly to the ChoreoMater network~\cite{chen2021choreomaster}, utilizing four GCN blocks and two Gated Recurrent Unit (GRU) layers to compress the dance sequence into 32-dimensional embedding vectors $Z_s$. The vectors are then input into an MLP classifier.
The dance style branch is pre-trained on the large-scale annotated dataset D2MIDI. In the final step, the beat binary sequence and style feature are concatenated to derive conditional control information $Z$, which subsequently guides the multi-instrument MIDI music generation.

\subsection{Music Representation}
\label{subsec: music_representation}
Inspired by SymphonyNet, we represent multi-instrument music using quads, which include event, duration, track, and instrument.
\begin{itemize}
    \item Event: The event attribute comprises four sub-attributes: measure, chord, position, and pitch. A BOM symbol is used to indicate the beginning of each measure, with all symbols in the measure added after the BOM symbol. Beat and note duration are adopted as time units to divide each measure and determine position. The pitch range is divided from 0 to 127 based on the general MIDI design.
    \item Duration: It represents the duration of each note.
    \item Track\&Instrument: The track and instrument attributes are determined by traversing the music and identifying the track and instrument corresponding to each note.
\end{itemize}
Unlike natural language sequences, symbolic music sequences exhibit relative position invariance. For instance, a chord $C$ containing the music notes $(C,N_1,N_2,N_3)$ is equivalent to $(C,N_2,N_1,N_3)$. As they comprise the same notes and are controlled by the same chord, the order of the notes does not affect the musical effect. Therefore, we employ relative position encoding for music notes.

\subsection{Drum Rhythm Generator}
Given that multi-instrument MIDI music can be represented as discrete tokens, it is inherently suitable for sequence modeling in the realm of natural language processing. Consequently, we employ an autoregressive approach to generate drum track notes. More specifically, we utilize a Transformer model~\cite{huang2018music} to generate drum notes in a step-by-step manner, guided by the dance condition control vector $Z$.
For the key attention module in the model, we adopt the Masked Self-Attention (MSA) module consistent with the Transformer and design the cross-attention module Video Guided MIDI (VGM). The cross-attention mechanism~\cite{chen2021crossvit} is utilized to blend two distinct sequences of embeddings, where these sequences can represent different modalities. Similarly, one sequence serves as the input query (Q), defining the length of the output sequence, while another sequence provides the input keys (K) and values (V).
The MSA and VGM modules are employed in pairs. The VGM module employs conditional control information $Z$ to guide attentional learning in the Drum Rhythm Generator. The attention maps of the VGM block tend to focus on values related to visual information. The specific calculation method is shown in Eq.~(\ref{equ:VGM}). Among them, the Drum-encoded sequence $D$ is utilized as the query, while the conditional control information $Z$ extracted from the dance video is employed as both the key and value. The parameter matrices $W^q \in \mathbb{R}^{d_{\text {model }} \times d_q}$, $W^k \in \mathbb{R}^{d_{\text {model }} \times d_k}$, and $W^v \in \mathbb{R}^{d_{\text {model }} \times d_k}$. In this work, the number of heads $h$ in the multi-head VGM attention module is 8. For each head, we use $d_q = d_k = d_{model}/h = 64$.
\begin{equation}
\operatorname{VGM}(Q,K,V)=\operatorname{softmax}\left(\frac{DW^q(ZW^k)^T}{\sqrt{d_k}}\right)(ZW^v) 
\label{equ:VGM}    
\end{equation}
\subsection{Multi-Track MIDI BERTGen}
In the process of enriching the entirety of a musical piece, we generate note information for tracks beyond the drum track. This task bears resemblance to image inpainting in computer vision and context understanding tasks in natural language processing. Given the unique nature of symbolic music as sequence modeling, we incorporate the BERT model to comprehend the entire symbolic music sequence. The audio track that is to be completed is considered a part of the random mask in the BERT model. Unlike the masking strategy used in the BERT model, we have designed a novel masking approach tailored to the characteristics of symbolic music composition. Considering that a piece of music comprises multiple measures, and each measure contains tokens with similar attributes – the values of signature, tempo, and measure attributes remain consistent within each measure, and the types of instruments also follow a similar pattern, restricted within a small-scale range. Within the same measure, the information on position and pitch is also closely related. Therefore, employing a masking strategy within the same measure facilitates the model's learning of musical pattern structures. Specifically, we apply masking to the same type of tokens (events, durations, tracks, instruments) across different measures. In alignment with BERT, we replace 80\% of all masked tokens with MASK tokens, substitute 10\% with a randomly chosen token, and leave the remaining 10\% unaltered.
The Multi-Track MIDI BERTGen, a classic multilayer bi-directional Transformer encoder, comprises 12 layers of multi-head self-attention, each with 12 heads, and a hidden space dimension of 768 in the self-attention layers. As a self-supervised method, BERTGen does not require labeled data from downstream tasks for pre-training. 

Each input token is initially transformed into a token embedding via an embedding layer, supplemented with a relative positional encoding that corresponds to its time step in the sequence. This is subsequently fed into a stack of 12 self-attention layers to obtain a contextualized representation, known as a hidden vector or hidden states, at the output of the self-attention stack. Owing to the bi-directional self-attention layers, the hidden vector is contextualized in that it has attended to information from all other tokens from the same sequence. Ultimately, the hidden vector of a masked token is fed into a dense layer to predict the missing token. As the vocabulary sizes for the four token types vary, we proportionally weight the training loss associated with tokens of different types to the corresponding vocabulary size to facilitate model training.
\subsection{Training and Inference}
Our model is trained in an end-to-end manner. During training, dance motion features and historical MIDI event sequences are input to predict the probability output of the next music event token. In the inference phase, the model autoregressively predicts the next MIDI event. Notably, at time step 0, the historical MIDI event sequence is empty, meaning that generation begins with an empty token.

\section{Experiments}
\label{subsec:experiments}

\subsection{Datasets}
We evaluated the effectiveness of our method through experiments on two datasets with paired dance video and music: the publicly available AIST dataset and our D2MIDI dataset. The non-repetitive music in the AIST dataset comprises only 60 pieces, with one piece of music corresponding to multiple dance segments. The AIST dataset contains a total of 1,618 dance motions. However, many motions are filmed from different camera perspectives, resulting in 13,940 dance videos. Thus, on average, one piece of music corresponds to 232 dance videos. The AIST dataset encompasses ten dance genres: ballet jazz, street jazz, krump, house, LA-style hip-hop, middle hip-hop, Waack, lock, pop, and break. In contrast, the music corresponding to each dance segment in our D2MIDI dataset is unique, making it more suitable for music generation tasks. The D2MIDI dataset contains 71,754 paired dance videos and MIDI music data, which are not present in the AIST dataset. The dance types of the D2MIDI dataset encompass six major genres:  classical, hip-hop, ballet, modern, latin, and house.
\begin{figure*}[t]
    \centering
    \includegraphics[width=1\linewidth]{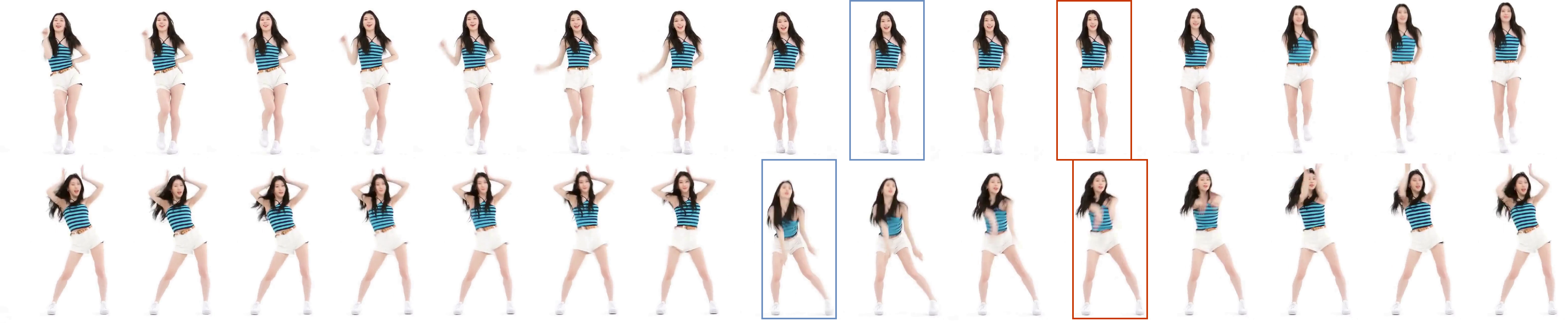}
    \caption{Visualization result. For the given dance video input, Dance2MIDI generates corresponding MIDI music and converts it into a waveform. The music beat is detected using the public toolbox Librosa. Two pieces of the dance videos are examples, where the blue box indicates the real dance beat (the turning point of the dance motion), and the red box indicates the frame of the dance video corresponding to the timestamp of our audio beat.}
    \label{fig:vis}
\end{figure*}
\subsection{Evaluation Metrics}
We evaluated our method both objectively and subjectively using publicly available metrics~\cite{di2021video,zhu2022quantized} and compared our model with the three State-of-the-Art models.

\subsubsection{Coherence}
To assess the coherence between dance beats and generated music rhythms, we utilize two objective metrics: Beats Coverage Score (BCS) and Beats Hit Score (BHS), as employed in previous works~\cite{Da2018Visual,lee2019dancing}.
These works have demonstrated that dance motions and music beats are typically aligned, allowing for a reasonable evaluation of music tempo by comparing beats in generated and ground truth music. We denote the number of detected beats in the generated music samples as $B_g$, the total number of beats in the original music as $B_t$, and the number of aligned beats in the generated samples as $B_a$. 
During inference, the duration of music generated using the different methods may not align precisely with the duration of the real music, resulting in an overflow of detected beats $B_g$. Consequently, when the duration of music generated by the model does not match or significantly differs from the ground truth, BCS may exceed its value range of [0,1]. To address this issue, we standardize BCS value and propose Beat Average Score (BAS) in conjunction with the BHS. Specifically, as shown in Eq.~(\ref{equ:BAS}), we employ exponential functions to constrain the BCS values within the range of 0 and 1. If the BCS value is less than 1, we use the exponential function $e^{\text{BCS}-1}$. When BCS is greater than 1, we apply the exponential function $e^{1-\text{BCS}}$.
\begin{itemize}
    \item \textbf{BCS}\quad It is calculated as $B_g/B_t$, representing the ratio of overall generated beats to total music beats. 
    \item \textbf{BHS}\quad It is calculated as $B_a/B_t$, representing the ratio of aligned beats to total musical beats.
    \item \textbf{BAS}\quad The calculation method is shown in Eq.~(\ref{equ:BAS}), representing the overall coherence between dance beats and music rhythms.
\end{itemize}
\begin{equation}
    BAS=\left\{\begin{array}{ll}
0.5 \times (e^{\text{BCS}-1}+\text{BHS}) & \text {s.t.} \text{BCS}<1 \\
0.5 \times (e^{1-\text{BCS}}+\text{BHS}) & \text {s.t.} \text{BCS}>1
    \label{equ:BAS}
\end{array}\right.
\end{equation}
\subsubsection{Quality}
We employed the objective metrics outlined in~\cite{di2021video,wu2020jazz} to assess the quality of symbolic music. These metrics include Pitch Class Histogram Entropy (PHE) and Grooving Pattern Similarity (GS). 

\textbf{PHE}\quad It evaluates the tonal quality of the music. we first gather all notes within each bar and then construct a 12-dimensional pitch class histogram $\overrightarrow{h}$, based on the pitch of all notes. This histogram is normalized by the total note count within the period such that $\sum_i h_i=1$. Then, we calculate the entropy of $\overrightarrow{h}$ as Eq.~(\ref{equ:phe}):

\begin{equation}
    \mathcal{H}(\overrightarrow{h})=-\sum_{i=0}^{11} n_i \log _2\left(h_i\right)
    \label{equ:phe}
\end{equation}

\textbf{GS}\quad It measures the rhythmicity of the music. In information theory~\cite{shannon1948mathematical}, entropy is utilized to measure uncertainty. We apply entropy $\mathcal{H}$ to assess the tonal quality of symbolic music. A lower entropy value $\mathcal{H}$ indicates a clear tonality within a piece of music. For GS, it represents the positions within a bar where at least one note onset occurs, denoted by $\overrightarrow{g}$, a 64-dimensional binary vector. The similarity between a pair of grooving patterns is defined as Eq.~(\ref{equ:gs}).

\begin{equation}
GS \left(\overrightarrow{\mathbf{g}}^a, \overrightarrow{\mathbf{g}}^b\right)=1-\frac{1}{Q} \sum_{i=0}^{Q-1} \operatorname{XOR}\left(g_i^a, g_i^b\right)
\label{equ:gs}
\end{equation}

Where $Q$ represents the dimension of $\overrightarrow{g}^a$ and $\overrightarrow{g}^b$,  $\operatorname{XOR}$ denotes the exclusive $\operatorname{OR}$ operation. If the music exhibits a distinct rhythmic feel, the groove pattern between each pair of bars should be similar, resulting in a high Groove Similarity score.
\begin{table*}[t]
\centering
\begin{tabular}{c|ccc|c|ccc|c}
\toprule
  & \multicolumn{4}{c|}{D2MIDI} & \multicolumn{4}{c}{AIST} \\ \cmidrule(lr){2-9}
Metric  & CMT  & Dance2Music  & D2M-GAN & ours & CMT  & Dance2Music & D2M-GAN       & ours          \\ \midrule
PHE $\uparrow$    & 2.49 & 2.24             & /             & \textbf{2.89}  & 2.55 & 2.26             & /             & \textbf{2.92} \\
GS $\uparrow$     & 0.62 & 0.98             & /             & \textbf{0.99}  & 0.64 & 0.98             & /             & \textbf{0.99} \\
BCS     & 5.11 & 1.75             & 0.68          & \textbf{0.73}           & 4.87 & 1.73             & 0.70          & \textbf{0.76}         \\
BHS     & 0.29 & 0.42             & 0.45          & \textbf{0.53}          & 0.32 & 0.44             & 0.48          & \textbf{0.61}          \\
BAS $\uparrow$ & 0.15 & 0.44          & 0.59 & \textbf{0.65}       & 0.17 & 0.46                & 0.61 & \textbf{0.69}         \\ \midrule
Consistency $\uparrow$    & 3.21 & 2.82             & 2.55          & \textbf{3.91}  & 3.38 & 2.99             & 2.62          & \textbf{3.99} \\
Noise $\uparrow$  & 3.43 & \textbf{3.68}    & 2.68          & 3.57           & 3.45 & \textbf{3.72}    & 2.82          & 3.67          \\
\bottomrule
\end{tabular}
\caption{Objective and subjective evaluation results on the D2MIDI and AIST Dataset. $\uparrow$ means the higher the better.}
\label{tab: D2MIDI and AIST}
\end{table*}
\subsubsection{Qualitative Evaluation}
We also conducted an audio-visual survey to subjectively compare the different models. We conducted the Mean Opinion
Score experiments~\cite{zhu2022quantized} to assess the quality of the music and the correspondence between the video and music. For each dance genre, 50 samples are evaluated by 5 professional choreographers. That is, a total of 500 evaluation samples are provided for the AIST dataset, and a total of 300 evaluation samples are provided for the D2MIDI dataset. 
Among the five choreographers involved in this study, three are female and two are male, spanning an age range of 25 to 45 years. They possess extensive experience in choreographing various types of dances. The entire evaluation process was conducted anonymously, ensuring that participants were unaware of which model generated the data samples.
In our study, we presented human participants with the same video accompanied by music synthesized using different methods. Participants are then asked to rate the music on a scale of 1 to 5, with higher scores indicating better performance. The primary evaluation criteria are: 
\begin{itemize}[leftmargin=3mm]
    \item \textbf{Consistency}\quad the degree to which the major stress or boundaries of the generated music aligned with the video boundaries or visual beat. For instance, fast-paced dance movements should be accompanied by major stress to enhance musicality.
    \item \textbf{Noise}\quad noise degree of sounds produced by non-instrumental sources. For instance, a pleasing musical composition should be free of extraneous white noise. The lower the noise level, the higher the score awarded by participants.
\end{itemize}
\subsection{Implement Details}
We applied the same processing method described in Sections~\ref{subsec:transcrip} and \ref{subsec:motion_estimation} to the AIST dataset to obtain paired dance motion joint data and MIDI music data. 
Our framework is implemented using PyTorch. The encoder of the graph convolutional network in our framework comprises 10 layers with residual connections.
The number of layers in the graph convolution network is set with reference to the ST-GCN network~\cite{yan2018spatial}, which is specifically designed for action recognition tasks. It consists of 9 layers of spatial-temporal graph convolution operators. The first three layers have an output of 64 channels. The following three layers have an output of 128 channels. The last three layers have an output of 256 channels. Subsequently, we add a layer with 512 channels to align with the dimension of the Transformer Encoder for predicting the beat binary sequence.
To prevent overfitting, we apply random affine transformations to the skeleton sequences of all frames during training as a data augmentation technique. Both the encoder and decoder blocks of the Drum Rhythm Generator are set to 6. For each block, the dimensionality of the attention layer and feed-forward network layer are set to 512 and 1024, respectively. The multi-head VGM attention block has 8 heads. For post-processing of the generated MIDI music data, we use the FluidSynth~\cite{newmarch2017fluidsynth} software synthesizer to convert the generated MIDI music into music waveform, consistent with the CMT model. We train our model using the Adam optimizer with parameters $\beta_1=0.9$, $\beta_2=0.9$, and $\varepsilon=10^{-9}$. The learning rate is scheduled during training with a warm-up period: it linearly increases to 0.0007 for the first 6000 training steps and then decreases proportionally to the inverse square root of the step number.

\subsection{Results}
We conducted training and evaluation on the AIST and D2MIDI datasets, respectively. For the state-of-the-art methods, including the CMT, Dance2\-Music, and D2M-GAN, we use the default parameter settings provided in their open-source code. For different dance genres, we divide the training, validation, and test sets in a ratio of 8:1:1 and use the same dataset settings for all models. As shown in Table~\ref{tab: D2MIDI and AIST}, our model outperforms existing state-of-the-art methods on the objective metrics of PHE and GS, indicating that the music we generated was slightly better in terms of tone and rhythm. In a similar vein, our method outperforms others on BHS, BCS, and BHS. This demonstrates that the congruence between the music and videos generated by our approach is superior. 
Our method also achieves the best performance on the subjective metric of Consistency, further indicating that the music and dance videos we generated are highly consistent.

Dance2Music achieves optimal performance on the Noise metric, a subjective indicator measuring the purity of music. This is because the Dance2Music model only models piano music and can only generate piano music, resulting in limited scalability. 
In the case of the CMT model, it has been observed that the length of the generated music often does not align with the duration of the corresponding dance. This discrepancy can be attributed to the model’s lack of explicit consideration for dance characteristics, resulting in a BCS metric that is significantly greater than 1. Our model is much stronger than that of the CMT model and the duration of the generated music is generally consistent with that of the dance video.

The visualization results are presented in Fig.~\ref{fig:vis}, where two dance videos are depicted. The red box represents the video frame corresponding to the timestamp of the generated music beat, while the blue box represents the real dance beat. It can be observed that there is a difference of only three video frames between them, indicating a high degree of alignment between the generated music and dance movements.

\section{Discussion}
In this paper, we constructed the first multi-instrument MIDI and dance paired dataset (D2MIDI), which can serve as a benchmark dataset for future research on generating background music for dance videos. We proposed the Dance2MIDI framework for multi-instrument MIDI generation. Dance2MIDI leverages the consistency of paired data to mitigate the weak correlation between music and video. 
As a two-stage generation framework, Dance2MIDI initially synthesizes a fundamental drum rhythm track utilizing the Transformer cross-attention mechanism, guided by dance condition information. Subsequently, the synthesis of the remaining audio tracks is structured as a sequence completion task. With the aid of the BERT model, we inpaint the remaining audio tracks.

In addition, some dance genres such as folk and ballet, may not exhibit the strong rhythmic elements characteristic of pop dance and often utilize music without drums. In the Dance2MIDI framework, we initially generate the drum track for the entire piece using the Drum Rhythm Generator module. This drum part, serving as the cornerstone of the music's beat, is enriched with the transitions and dynamics of dance movements and assists in the generation of other tracks within the Multi-Track MIDI BERTGen module. For dances like folk and ballet, we opt to remove the drum track in the final stage through post-processing, resulting in music without drums. Adopting this pipeline enhances generalizability, making it applicable to various dance types.

However, there are still limitations to our work: due to the variability in shape, form, and mechanics of drum instruments~\cite{su2021does}, their performance is a major bottleneck for music quality, which we aim to address in future work. 
Additionally, Dance2MIDI is currently limited to the generation of soundtracks for single-person dances. Future work will include an in-depth exploration of the application of group dances. 
For the user study, the current qualitative assessment experiment is indeed limited by the number of participants.
We plan to extend our coverage to diverse participant groups in the future. This will be achieved by randomly selecting participants, a method aimed at minimizing self-selection bias.

\bibliographystyle{IEEEbib}
\bibliography{refs}

\end{document}